\documentclass[lettersize,journal]{IEEEtran}
\usepackage{algorithm}
\usepackage{algorithmic}
\usepackage{amsfonts}
\usepackage{array}
\usepackage[caption=false,font=normalsize,labelfont=sf,textfont=sf]{subfig}
\usepackage{textcomp}
\usepackage{verbatim}
\usepackage{color}
\usepackage{graphics}
\usepackage{epsfig}
\usepackage{cite}
\usepackage{threeparttable}
\usepackage{graphicx}
\usepackage{picinpar}
\usepackage{amssymb}
\usepackage[cmex10]{amsmath}
\usepackage{subfloat}
\usepackage{stfloats}

\usepackage{bm}
\usepackage{url}
\usepackage{booktabs}
\usepackage{setspace}
\usepackage{multicol}
\usepackage{multirow}

\input epsf

\begin{document}

\title{\huge{Vision-Intelligence-Enabled Beam Tracking for Cross-Interface Optical Wireless Communication between Underwater and Low-Altitude Platforms}}


\author{ 
{Jiayue Liu, Tianqi Mao,~\IEEEmembership{Member,~IEEE}, Leyu Cao, Weijie Liu, Dezhi Zheng, Julian Cheng,~\IEEEmembership{Fellow,~IEEE}, \\ and Zhaocheng Wang,~\IEEEmembership{Fellow,~IEEE}
}

\thanks{J. Liu, T. Mao, L. Cao, and D. Zheng are with the State Key Laboratory of Environment Characteristics and Effects for Near-space, Beijing Institute of Technology, Beijing 100081, China (e-mail: jiayue\_liu@bit.edu.cn, maotq@bit.edu.cn, leyu\_cao@bit.edu.cn, zhengdezhi@bit.edu.cn).}
\thanks{W. Liu is with CAS Key Laboratory of Wireless-Optical Communications, School of Information Science and Technology University of Science and Technology of China, Hefei, China (e-mails: lwj1993@ustc.edu.cn).}
\thanks{J. Cheng is with the School of Computing and Information Technology,  Dongguan, Guangdong, China (email: jcheng@gbu.edu.cn) and he is also with the School of Engineering, The University of British Columbia, Kelowna, BC, Canada (email: julian.cheng@ubc.ca).
}
\thanks{Z. Wang is with Department of Electronic Engineering Beijing National Research Center for Information Science and Technology Tsinghua University, Beijing, China (e-mail: zcwang@tsinghua.edu.cn).}
}
\maketitle

\begin{abstract}
The rapid expansion of oceanic applications such as underwater surveillance and mineral exploration is driving the need for real-time wireless backhaul of massive observational data. Such demands are challenging to meet using the narrowband acoustic approach. Alternatively, with the participation of low-altitude platforms (LAPs), water-air optical wireless communication (OWC) has emerged as a promising solution owing to its high potential for broadband transmission. However, implementing water–air OWC remains challenging, particularly when signals penetrate the fluctuating interface, where dynamic refraction induces severe beam misalignment with airborne stations. This necessitates real-time transceiver alignment capable of adapting to complex oceanic dynamics, which remains largely unaddressed. Against this background, this paper establishes a mathematical channel model for water–air optical transmission across a time-varying sea surface. Based on the model, a vision-based beam tracking algorithm combining convolutional neural network and bi-directional long short-term memory with an attention mechanism is developed to extract key spatio-temporal features. Simulations verify that the proposed algorithm outperforms classical methods in maintaining received signal strength and suppressing vision noise, demonstrating its robustness for water–air OWC systems.

\end{abstract}


\vspace{-3mm}

\section{Introduction}\label{s1}

The escalating development of maritime engineering has driven unprecedented requirement on data transmission under oceanic scenario. 
To satisfy progressively diversified needs, communication links involving low-altitude and underwater unmanned platforms provide the most flexible and effective terminal structure in marine communication networks.
In contemporary scenarios, underwater platforms such as unmanned underwater vehicles (UUVs) are extensively used for diverse marine exploration tasks, necessitating underwater data transmission support\cite{An_overview}. 
Low-altitude platforms (LAPs) such as unmanned aerial vehicles (UAVs) serve as a critical structure in the network that keep connection with UUVs and maintain communication backhaul. 
These applications have further motivated the implementation of water-air transmission that enables cooperation between various equipment and facilities for complicated tasks\cite{Real_time}.
In response to ever-increasing demands, researchers have developed various theoretical frameworks and technical methodologies for oceanic communication systems.
In recent research, optical wireless communication (OWC) has emerged as a replacement for the traditional acoustic approaches to oceanic communication requests, offering high transmission rates, reliable security, and an acceptable propagation range\cite{A_Survey_of_Underwater,Underwater_and}. 
The OWC cross-interface technology has obviously greater potential in certain applications, attracting various research efforts toward the area.
Early research efforts have demonstrated successful underwater-to-air communication through two primary approaches: relay-based communication systems and direct trans-interface optical wireless communication\cite{Recent_Progress}.
\cite{Dual_hop} and \cite{A_software} proposed relay-based systems which contain multi-hop transmission and integrated acoustic and optical network. 
Moreover, recent research shows the direct communication approaches require none additional equipment\cite{Real_time,Multi_user,Mitigation}.
Given its low system complexity, flexible deployment, and enhanced security, direct cross-interface communication is highly desirable. 

Despite the advantages of water-air direct cross-interface OWC, there still exist numerous challenges to its practical implementation. 
In addition to propagation attenuation, absorption and scattering losses\cite{Waving_Effect}, and noise interference\cite{Background_Noise}, the primary challenge faced by water-air OWC is beam misalignment resulting from optical path perturbations caused by the dynamic interface\cite{Underwater_and}. 
The effects of refraction across the interface and dynamics of the oceanic surface influence the optical propagation trajectory, subsequently affecting the line-of-sight path between UUV and UAV, leading to reduction of signal reception intensity and stability.
To address this issue, various studies have implemented diverse methodologies to enhance the quality of cross-interface communication systems.
Wide-beam LED transmitters have emerged as a straightforward solution, avoiding misalignment by its extensive spatial coverage of the optical beam\cite{Field_Demonstrations,Preliminary}. 
Further research implementing multi-input multi-output techniques to enhance transmission rate while supporting multi-unit transmission\cite{Real_time}\cite{Multi_user}. 
Beam alignment has emerged as a widely acknowledged method for this challenge, which facilitates the maintenance of high-intensity optical communication path through dynamic adjustment of transceiver orientations in response to water-air interface variations.
A method combining retro-reflector and photodiode (PD) array is applied for sensing the variation of the dynamic interface, while establishing a direct tracking link to compensate for the misalignment of the optical beam due to negative feedback\cite{Mitigation}. 
In addition, further research incorporates micro-electro-mechanical systems (MEMS) combined with PD array feedback to achieve more precise beam alignment leading to an increase in receiving intensity\cite{Real-Time_Wave}. 
With the advancement of artificial intelligence (AI), algorithms that address complex problems through artificial neural networks have been widely adopted. 
A beam tracking method aided by vision AI based on 2-dimensional convolutional neural network (CNN) has made progress in the water-air attenuation channel\cite{BeamTrackingAided2024}. 
Even though the CNN is used only for preprocessing to improve signal quality, it motivates applying neural networks to extract information reliably in challenging environments.
Moreover, a transformer-based camera demodulator has been proposed, which recognizes and tracks the cross-interface optical signal by software processing\cite{Bubble}.
Unsupervised learning has also been explored for this scenario. For example, deep reinforcement learning (DRL) based beam alignment approaches have been proposed \cite{Reinforcement_learning,Enabling_Reliable}, which move beyond traditional methods and enable passive beam tracking.

Based on the above research, existing water-air OWC faces the following challenges: a). In sophisticated environments such as oceans, dynamic changes at cross-medium interfaces affect direct signal transmission. b). Traditional methods cannot effectively sense and predict the dynamics of the sea surface for adaptive adjustments. c). AI methods with better performance have not been systematically applied to cross-interface communication technologies.
To address these issues, this paper proposes a vision-intelligence-enabled beam tracking algorithm to enhance the performance of water-air cross-interface OWC. 
The main contributions of our work are highlighted as follows:
\begin{itemize}
    \item We establish a water-air interface refraction model based on wave spectrum theory. Furthermore, we demonstrate the uniqueness of the optical path across the interface using minimum OPL theory, thereby enabling the modeling process of a water-air cross-interface optical wireless communication channel that incorporates the refraction and attenuation model.
    \item We propose an algorithm based on a ray-tracing method to construct the vision model for directional information sensing. Visual images consist of discrete pixels, where each pixel intensity is computed using the RT algorithm by accumulating light intensity contributions from both light sources and refractive interfaces during the back-tracking process.
    \item We propose a beam tracking algorithm based on deep learning. CNN and bi-LSTM structure are applied to extract information over vision data and time-relevant sequence. Moreover, attention mechanism is used to increase the awareness of the critical directional information. The proposed algorithm demonstrates better performance compared to its counterparts under complicated environments.
\end{itemize}


\section{System Model}\label{s2}


\begin{figure}[!t]
    \centering
    \includegraphics[width=0.95\linewidth, keepaspectratio]{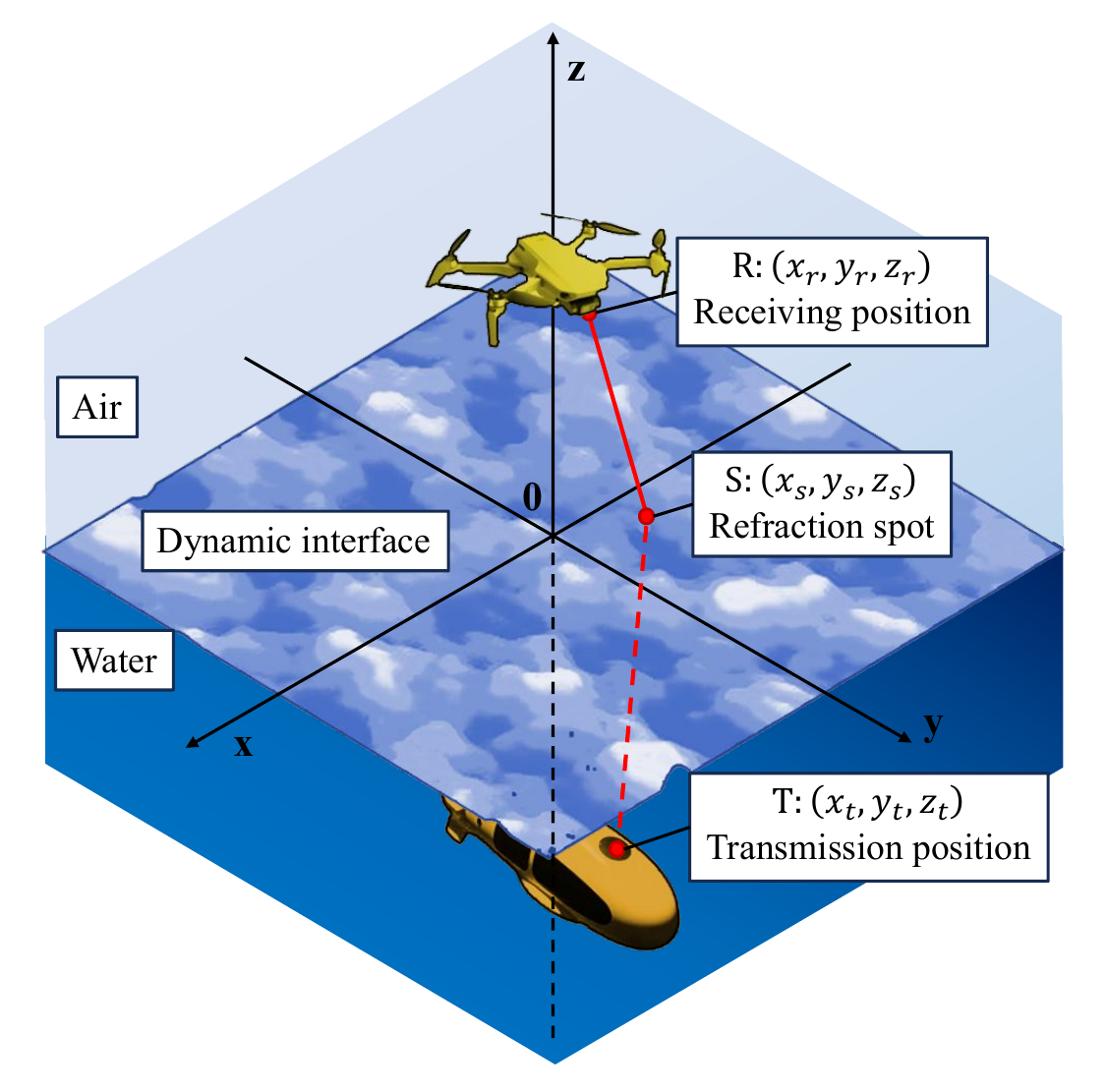}
    \caption{OWC channel Model between underwater platform and low-altitude platform.}
    \label{sysmodel}
    \vspace{-3mm}
\end{figure}

Consider a cross-interface water-air OWC system in 3-dimensional space, which is illustrated in Fig. \ref{sysmodel}. 
Specifically, the optical signal is emitted by laser diode (LD) at position $\mathrm{T}:(x_t,y_t,z_t)$, crossing the water-air interface at position $\mathrm{S}:(x_s,y_s,z_s)$, and received by avalanche photodiode (APD) at $\mathrm{R}:(x_r,y_r,z_r)$.
Note that the beam emitted from the LD exhibit strong directionality, making transceiver alignment extremely crucial for stable communication link\cite{High_speed_645}. 
To facilitate relevant theoretical analysis/innovations, below we establish a mathematical model for cross-interface optical transmission, which clearly characterizes the deleterious impacts on beam alignment from the fluctuating sea surface.

\subsection{Propagation Path of the Optical Beam}\label{s2.1}
Investigation of the optical propagation path can be mandatory for highly directional transmission, especially under the sophisticated water-air channel involving the dynamic behaviors of the sea surface.


\subsubsection{Ocean Wave Model}\label{2.1.1}

\begin{figure*}[!t]
    \centering
    \subfloat[]{
        \includegraphics[width=0.45\linewidth, keepaspectratio]{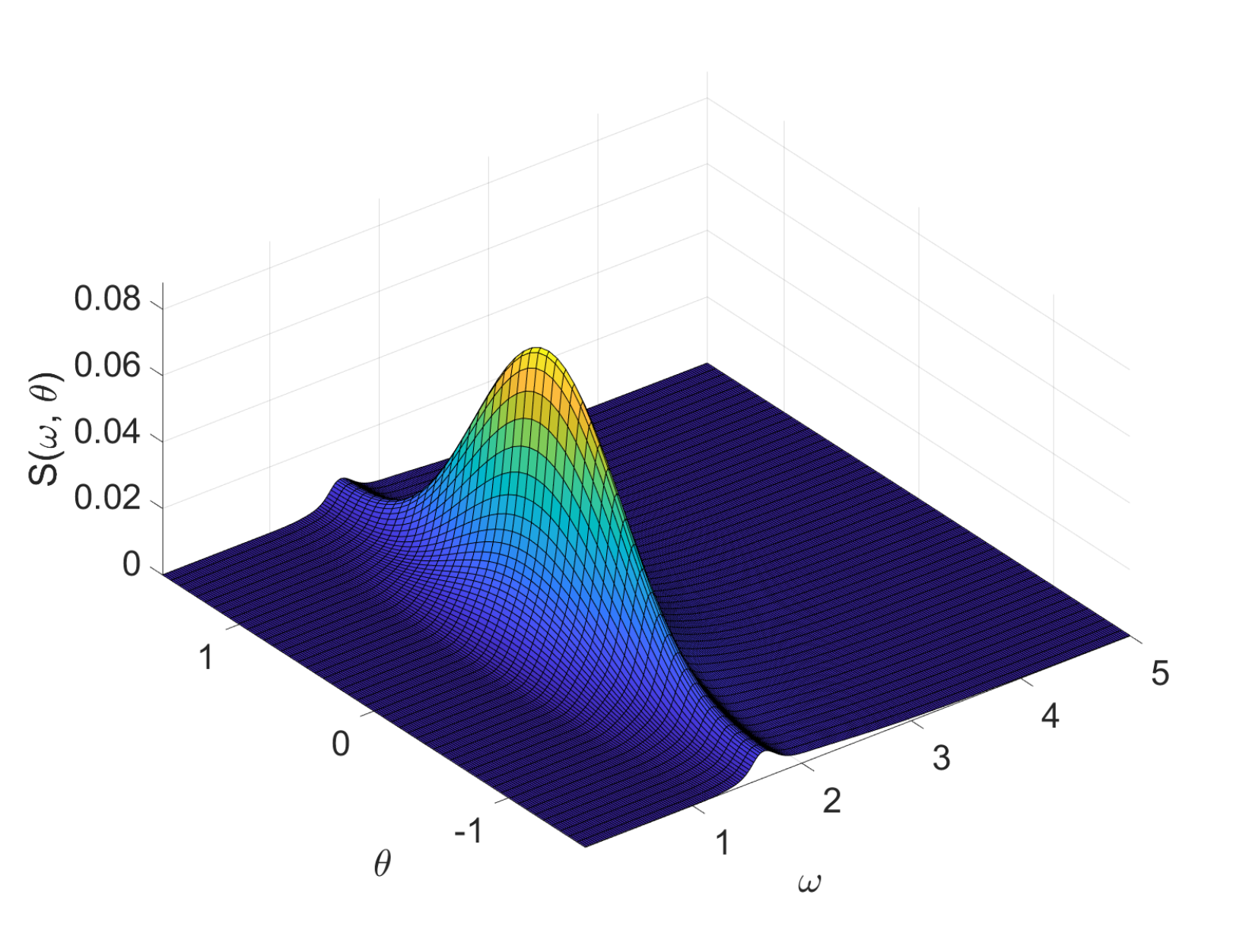}
    }
    \subfloat[]{
        \includegraphics[width=0.45\linewidth, keepaspectratio]{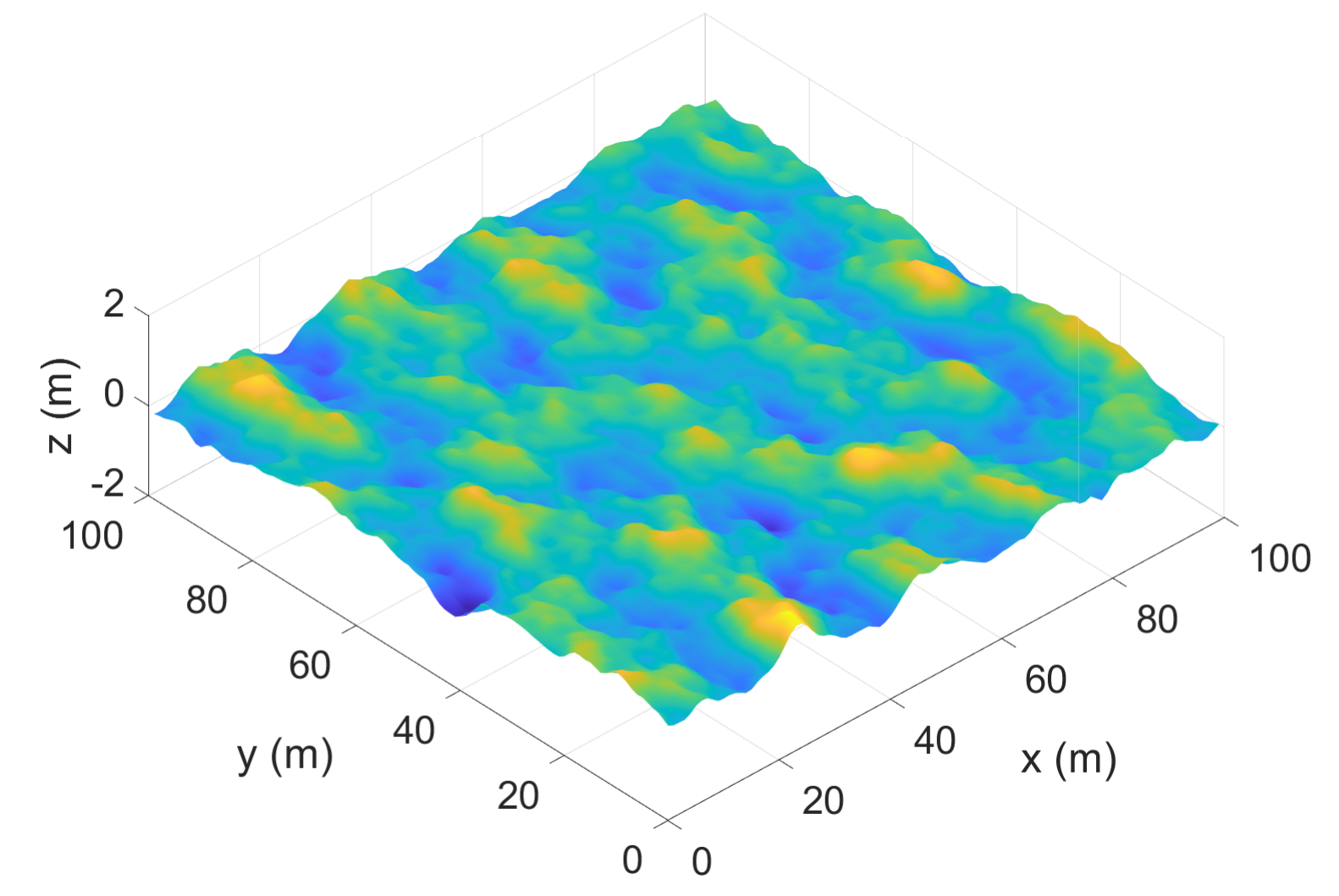}
    }
    \caption{Oceanic surface illustrations based on the wave spectrum theory, including: (a) 3-dimensional ocean wave spectrum. (b) Wave surface simulated by harmonic wave method.}
    \label{oceanwave}
    \vspace{-3mm}
\end{figure*}

The fluctuations of the sea surface, i.e., ocean waves, requires careful modeling to determine the propagation path across the interface\cite{Waving_Effect}.
Inspired by the ocean-wave spectrum theory in oceanography\cite{Three-dimensional}, we start from the ocean-wave spectrum $S(\omega,\theta)$, which is expressed as:
\begin{multline}\label{WaveSpectrum}
    S(\omega,\theta) = \frac{ag^2}{\omega^5}\exp[-1.25(\frac{\omega_p}{\omega})^4]·\gamma^{\exp[-\frac{(\omega-\omega_p)^2}{2(\sigma\omega_p)^2}]} \\ 
    \times \frac{1}{\pi}[1+p\cos(2\theta)+q\cos(4\theta)], \theta\le\frac{\pi}{2},
\end{multline}
where $\omega$ and $\theta$ stand for the angular frequency and directional angle of the harmonic wave components, respectively. $\omega_\mathrm{p}$ denotes the peak-power-distribution frequency, which is determined by the oceanic wind speed, wind fetch, and gravity. $p$ and $q$ are harmonic factors that are related to $e^{(\omega/\omega_\mathrm{p})^4}$. 
The introduced $S(\omega,\theta)$ describes the power distribution of the oceanic wave, which is shown as Fig. \ref{oceanwave} (a).
Based on the ocean-wave spectrum, the shape function of water-air interface, denoted by $W(x,y,t)$, can be constructed using the harmonic-wave method, formulated as\cite{Three-dimensional}
\begin{multline}\label{Harmonic}
    W(x,y,t)=\sum_i \sum_j \sqrt{S(\omega_i,\theta_j) d\omega d\theta} \\ \times\cos\left[ \omega_i t - \frac{\omega_i^2}{g}(x\cos\theta_j+y\sin\theta_j)+\epsilon_{i,j} \right],
\end{multline}
which describes the geometric information of the water-air interface, visualized as Fig. \ref{oceanwave} (b). In \ref{WaveSpectrum}, $S(\cdot)$ represents the ocean-wave spectrum, $\omega_i$ and $\theta_j$ represent frequency and direction angle, respectively, and $\epsilon_{i,j}$ stands for the random phase shift. 
Based on this shape function of $W(x,y,t)$, the beam propagation trajectory can be derived through the minimum optical path theory, which will be elaborated in the following section.

\subsubsection{Path of the Optical Signal}\label{2.1.2}
In this paper, we mainly consider direct propagation between the transceiver, where the scattering paths are neglected due to sever signal attenuation\cite{Research_on_Underwater}. 
For brevity, the direct path can be expressed by the line segments between $\mathrm{T}$, $\mathrm{S}$, and $\mathrm{R}$, illustrated in Fig. \ref{Channel_model}.
The direct path follows the theory of the minimum optical path length (OPL) that the optical beam should propagate along the path with the shortest equivalent length, which is formulated as $\mathrm{OPL} = n_1d_w + n_2d_a$\cite{Enabling_Reliable}, 
where $n_1$ and $n_2$ represent the refraction indices of water and air, $d_w=\| \overrightarrow{\mathrm{T}\mathrm{S}} \|$ represents the propagation distance in water, and $d_a=\| \overrightarrow{\mathrm{S} \mathrm{R}} \|$ denotes the propagation distance in air.
Obviously, when both transmission position $\mathrm{T}$ and receiving position $\mathrm{R}$ are fixed, the refraction point of the direct path can be obtained by minimizing the OPL under the constraint of surface shape function.
The optimization problem can be formulated as follows\cite{Enabling_Reliable}
\begin{equation}\label{minOPL}
\begin{split}
    & \underset{\mathrm{S}}{\mathbf{min}} \quad n_1d_w + n_2d_a \\
    & \ \mathbf{s.t.} \quad \mathrm{S} \in \{(x,y,z)|z=W(x,y,t)\}.
\end{split}
\end{equation}
The solution to (\ref{minOPL}) provides a spatial trajectory of the direct path determined by the coordinates of the transceiver and the dynamic interface geometry, which is the foundation for the modeling of the channel attenuation in the following.

\subsection{Path Gain Model}\label{2.2}
The overall path-gain coefficient is mainly determined by the transceiver gains and the propagation attenuation, which are elaborated below.
\subsubsection{Directional Attenuation at Transceiver}\label{s2.2.1}
In the optical transmitter, an LD serves as the signal-emitting device, whose optical characteristics lead to attenuation related to the angle of departure $\alpha_\mathrm{D}$\cite{Underwater_laser_C}.
The transmission intensity of the optical beam can be elaborated by a normalized function, which is expressed as:
\begin{equation}\label{departureGain}
    G_\mathrm{D}(\alpha_\mathrm{D},\lambda)=\exp(-\frac{2\sin^2\alpha_\mathrm{D}}{\omega_\mathrm{D}^2[1+(\lambda \cos\alpha_\mathrm{D}/(\pi\omega_\mathrm{D}^2))^2]}),
\end{equation}
where $\alpha_\mathrm{D}$ represents the angle of departure, $\lambda$ represents wavelength, and $\omega_\mathrm{D}$ represents the maximum emitting angle of the LD\cite{Improvement_of}.
In the optical receiver, APD acts as the signal detection device, whose optical characteristics lead to attenuation related to the angle of arrival $\alpha_\mathrm{A}$\cite{Study_on}.
The receiver attenuation is described by a normalized $G_\mathrm{A}(\alpha_\mathrm{A})$, expressed as:
\begin{equation}\label{arrivalGain}
    G_\mathrm{A}(\alpha_\mathrm{A})=\frac{n_2^2\cos\alpha_\mathrm{A}}{\sin^2\omega_\mathrm{A}},
\end{equation}
where $\alpha_\mathrm{A}$ represents the angle of arrival, and $n_2$ represents the refractive index of the air.
In conclusion, the intensity attenuation at transceiver can be quantized using angular-related equations, where the channel attenuation deteriorates with the increment of angular biases at the transceiver.

\subsubsection{Propagation Path Attenuation}\label{s2.2.2}
Path attenuation comes from two aspects: distance attenuation and media attenuation.
The distance attenuation results from the geometric separation of the optical beam, which is related to the inverse square of the overall propagation distance. 
Additionally, the media attenuation is caused by absorption and scattering of molecules and fragments in the medium, and follows the Beer-Lambert law\cite{Absorption_spec}.
The path attenuation in the media can be expressed as follow: 
\begin{equation}
    G_\mathrm{path}
     = \frac{\exp[-(a_w+b_w)d_{w}-(a_a+b_a)d_a]}{(d_{w}+d_{a})^2},
\end{equation}
where $d_w$ and $d_a$ represent the propagation distance in water and air medium, respectively. Moreover, $a_w(\lambda)$, $b_w(\lambda)$,$a_a(\lambda)$ and $b_a(\lambda)$ denote the underwater absorption coefficient, underwater scattering coefficient, aerial absorption coefficient, aerial scattering coefficient scattering coefficient, respectively.

\subsubsection{Refraction Attenuation across the Interface}\label{s2.2.3}
The dynamic interface of the water-air communication scenario causes unique refraction attenuation of the optical signal.
The energy of the refracted optical beam will be split by the reflection part when crossing the interface, which causes intensity reduction that can be derived according to the Fresnel's equation, formulated as \cite{Fresnel},
\begin{multline}\label{Fresnel}
    I_1 = I_0\bigg (1-\frac{1}{2}\bigg[(\frac{n_2\cos\theta_1-n_1\cos\theta_2}{n_2\cos\theta_1+n_1\cos\theta_2})^2     
    \\
    + (\frac{n_1\cos\theta_1-n_2\cos\theta_2}{n_1\cos\theta_1+n_2\cos\theta_2})^2\bigg] \bigg ),
\end{multline}
where $n_1$, $n_2$, $\theta_1$, $\theta_2$ represent the refractive index of water and air, the incident angle, and the refraction angle, respectively; $I_0$ and $I_1$ represent the intensity of incident light and refraction light, respectively. While the optical beam traverses through medium interfaces, the intensity attenuation caused by refraction is expressed as $G_\mathrm{ref} = I_1/I_0$, whose value can be calculated by (\ref{Fresnel}).

%
\begin{figure*}[!t]
    \centering
    \includegraphics[width=1\linewidth, keepaspectratio]{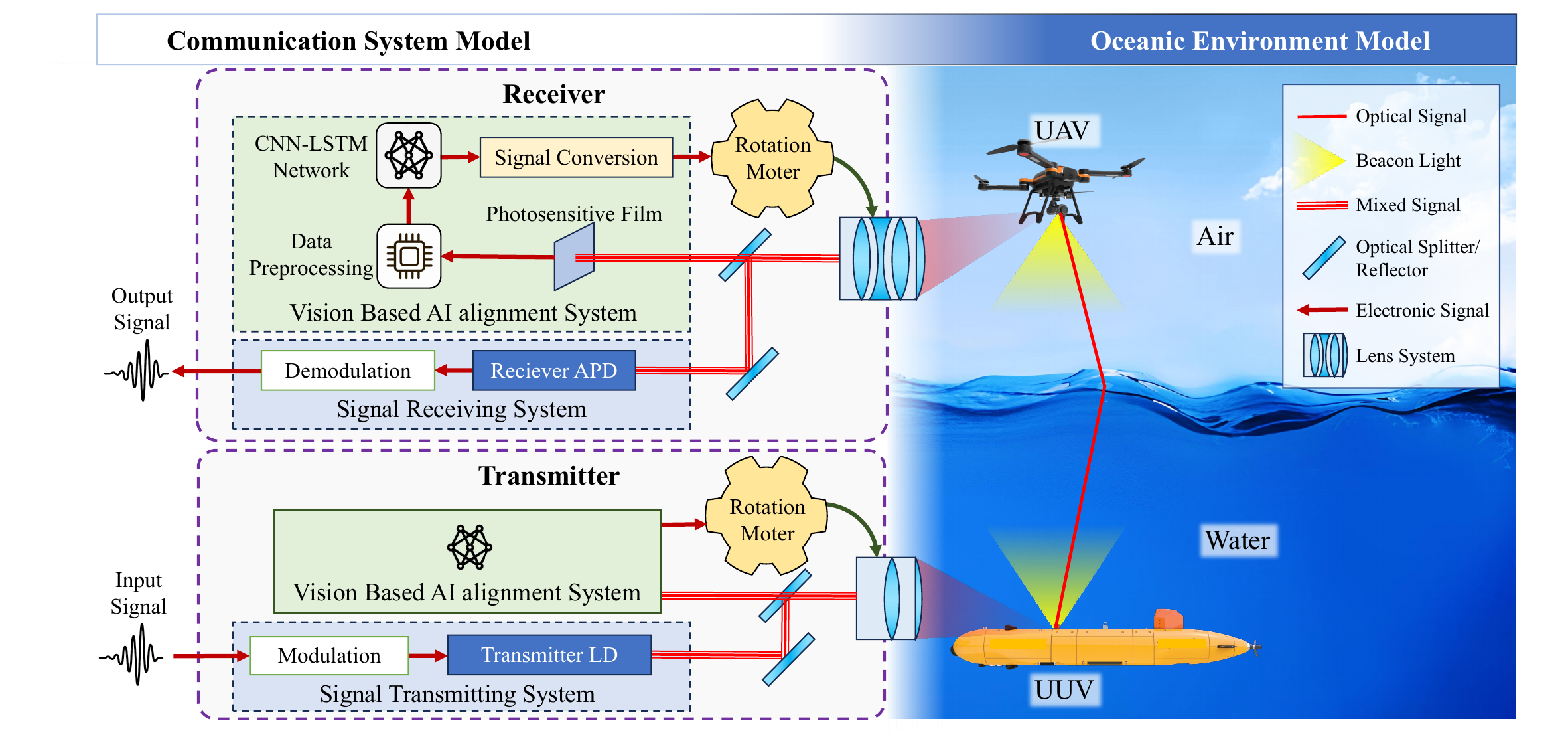}
    \caption{System model of the water-to-air OWC system within oceanic environment.}
    \label{Channel_model}
\end{figure*}


Based on the aforementioned, the overall channel gain coefficient can be calculated by combining the different factors above, which is expressed as:
\begin{equation}\label{Attenuation}
    G(\alpha_\mathrm{D},\alpha_\mathrm{A}) = G_\mathrm{D}(\alpha_\mathrm{D},\lambda)
    G_\mathrm{A}(\alpha_\mathrm{A})
    G_\mathrm{path}
    G_{\mathrm{ref}}.
\end{equation}

\subsection{Vision System Model}\label{s2.3}
\begin{figure*}[!t]
    \centering
    \includegraphics[width=0.95\linewidth, keepaspectratio]{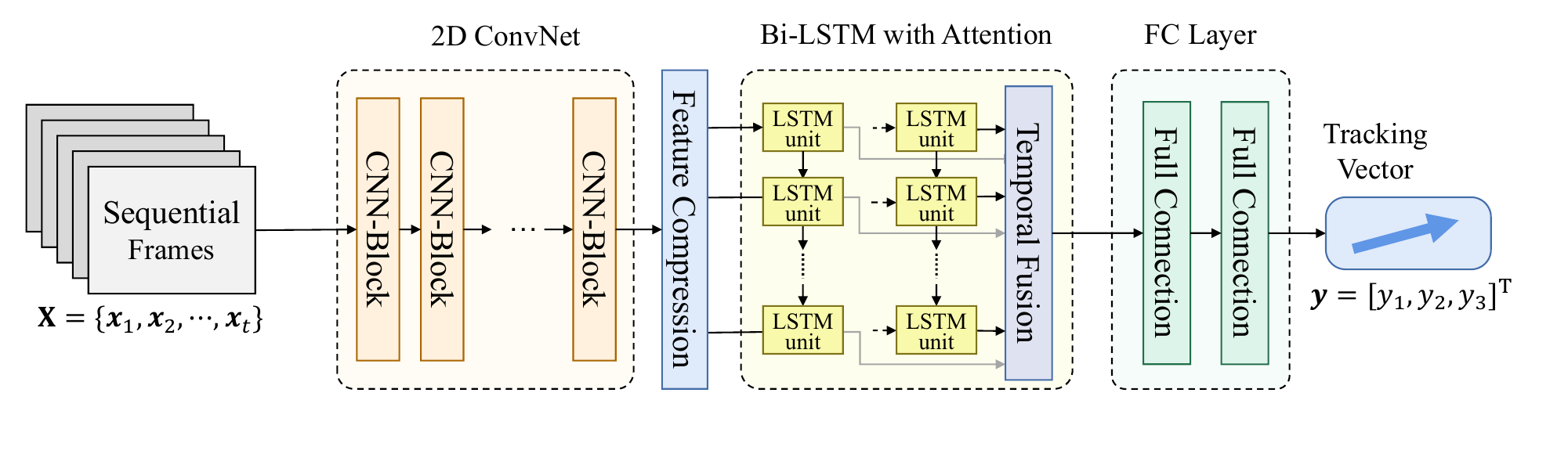}
    \vspace{-3mm}
    \caption{Model structure of the proposed AI Network.}
    \label{CNNLSTM}
\end{figure*}
\begin{figure*}[!b]
    \begin{multline}\label{networkFunc.}
        \mathbf{y} = \Theta(\mathbf{x}_t) = \tanh\Bigg (W_o\Bigg (\Theta_{\text{FC}}\Bigg[v^\top\tanh\bigg(W_a\left[\overrightarrow{\Theta_{\text{LSTM}}}\left(\Theta_{\text{CNN}}(\mathbf{x}_t), \boldsymbol{h}_{t-1},\boldsymbol{c}_{t-1}\right); \overleftarrow{\Theta_{\text{LSTM}}}\left(\Theta_{\text{CNN}}(\mathbf{x}_t), \boldsymbol{h}_{t+1}, \boldsymbol{c}_{t+1}\right)\right] + b_a\bigg)\Bigg]\Bigg ) + b_o\Bigg ),
    \end{multline}
\end{figure*}

As shown in Fig. \ref{Channel_model}, a photosensitive screen is equipped at both the transmitter and the receiver, which is employed to capture the beacon light and guide beam tracking process.
To simulate the vision data perceived by the screen, a ray-tracing (RT) algorithm is adopted to extract vision data from the simulation environment, as summarized in Algorithm 1 \cite{Ray_tracing}.
    

\begin{algorithm}[!t]
\caption{Ray Tracing for Data Acquisition}
\label{alg1}
\begin{algorithmic}[1]

\REQUIRE Position of the receiver $\mathrm{R}:(x_r,y_r,z_r)$, Direction of the receiver $\boldsymbol{r}_r$, position of the transmitter $T:(x_t,y_t,z_t)$, and the wave interface $W(x,y,t)$. 
\STATE $\mathbf{Initialize:}$ Intensity matrix $\mathbf{I} \in \mathbb{R^*}^{\mathrm{M}\times\mathrm{N}}$, focus position $\mathrm{F} = \mathrm{R} - f\boldsymbol{r}_r$, and pixel index $(i,j)\ \leftarrow\ (1, 1),\cdots,(M, N)$
\REPEAT
\STATE  \begin{minipage}[t]{\dimexpr\linewidth-\algorithmicindent-0.2cm}{\strut Initialize traceback ray by its initial position $\mathrm{B^{(0)}}$, and initial directional vector $\boldsymbol{r}^{(0)}$ according to (\ref{traceinit}); \strut}\end{minipage}
\STATE  \begin{minipage}[t]{\dimexpr\linewidth-\algorithmicindent-0.2cm}{\strut Trace back to the water-air interface by solving its position $\mathrm{B}^{(1)}$ based on (\ref{refractionSpots}), and directional vector $\boldsymbol{r}^{(1)}$ base on  (\ref{refractionVector}); \strut}\end{minipage}
\STATE  \begin{minipage}[t]{\dimexpr\linewidth-\algorithmicindent-0.2cm}{\strut Trace back to the water and judge whether the optical beam passes through the light source by (\ref{condition}); \strut}\end{minipage}
\STATE  \begin{minipage}[t]{\dimexpr\linewidth-\algorithmicindent-0.2cm}{\strut If (\ref{condition}) is satisfied, compute the intensity $\mathbf{I}(i,j)=\mathrm{I_0}G(\alpha_\mathrm{D}, \alpha_\mathrm{A})$ by (\ref{Attenuation}), otherwise, $\mathbf{I}(i,j)=0$;\strut}\end{minipage}
\STATE  \begin{minipage}[t]{\dimexpr\linewidth-\algorithmicindent-0.2cm}{\strut Update pixel index $(i,j)$;\strut}\end{minipage}
\UNTIL{The index $(i,j)$ reach $(M,N)$}
\RETURN Intensity pattern $\mathbf{I}$.
\end{algorithmic}
\end{algorithm}
Here, the objective of RT is to acquire a two-dimensional positive real vector $\mathbf{I} \in \mathbb{R^*}^{\mathrm{M}\times\mathrm{N}}$ that contains $M\times N$ intensity information pixels. According to the imaging principle, each backtracking ray passes through the focus point, which can be expressed as $\mathrm{F} = \mathrm{R} - f\boldsymbol{r}_r$\cite{Imaging}.
The algorithm traverses each pixel marked $(i,j)$ where $i=1,\ldots,M$ and $j=1,\ldots,N$.
In each iteration, a virtual ray is generated which traces the optical trajectory back from its pixel position $\mathrm{B}^{(0)}$, and along the direction of the vector $\boldsymbol{r}^{(0)}$ from the focus point to the pixel position, expressed as:
\begin{equation}\label{traceinit}
    \begin{aligned}
        & \mathrm{B}^{(0)} = \mathrm{R} + (i-M/2)d_\mathrm{px}\hat{\boldsymbol{e}_i} + (j-N/2)d_\mathrm{px} \hat{\boldsymbol{e}_j}, \\
        & \boldsymbol{r}^{(0)} = \overrightarrow{\mathrm{FB}^{(0)}},
    \end{aligned}
\end{equation}
where $d_\mathrm{px}$ denotes the interval of the adjacent pixels.
The traced ray intersects the water-air interface at position $\mathrm{B}^{(1)}:(x,y,z)$, where $z = W(x, y, t) = \mathrm{B}^{(0)} + k \cdot\boldsymbol{r}^{(0)}\ (k\in \mathbb{R}^*)$, and also changes direction according to the refraction effect across the interface. Thus, the tracing position and direction need to be renewed as follows: 
Specifically, when $W(x, y, t) = \mathrm{B}^{(0)} + k \cdot\boldsymbol{r}^{(0)}\ (k\in \mathbb{R}^*)$ is satisfied, the trace-back ray contact with the media interface, where the position and direction of the refraction point are expressed as:
\begin{multline}\label{refractionSpots}
    \mathrm{B}^{(1)} = (x,y,W(x,y,t)) |_{W(x, y, t) = \mathrm{B}^{(0)} + k \cdot\boldsymbol{r}^{(0)}}  ,
\end{multline}
\begin{multline}\label{refractionVector}
    \boldsymbol{r}^{(1)} = \frac{n_1}{n_2}(\boldsymbol{r}^{(0)} - (\boldsymbol{r} \cdot \hat{\boldsymbol{n}})\hat{\boldsymbol{n}}) 
    \\- \hat{\boldsymbol{n}}\sqrt{1 - (\frac{n_1}{n_2})^2(1 - (\boldsymbol{r}^{(0)} \cdot \hat{\boldsymbol{n}})^2)},
\end{multline}
where $W(x, y, t)$ represents the water-air interface function in (\ref{WaveSpectrum}), $n_1$ and $n_2$ represent the refractive index of air and water, and $\hat{\boldsymbol{n}}$ represents the normal vector of the water-air interface function that can be acquired by $\hat{\boldsymbol{n}}=[\frac{\partial W}{\partial x}, \frac{\partial W}{\partial y}, -1]^\top$.

In the next trace-back step, whether the ray passes through the light source determines the existence of the receiving intensity at the pixel. Assumed that the LD transmitter as a light source with a scale of $R_0$, located at $\mathrm{T}:(x_t,y_t,z_t)$, the condition that the ray passes through the light source can be judged as follow:
\begin{equation}\label{condition}
    \bigg\lvert\boldsymbol{r}^{(1)}\big|\overrightarrow{\mathrm{B}^{(1)}
    \mathrm{T}}\big| - \overrightarrow{\mathrm{B}^{(1)}\mathrm{T}}\bigg\rvert \leq R_0.
\end{equation}
If the condition expressed in (\ref{condition}) is satisfied, the intensity of pixel $\mathbf{I}(i,j)$ can be calculated by $\mathbf{I}(i,j)=\mathrm{I_0}G(\alpha_\mathrm{D}, \alpha_\mathrm{A})$, where $\mathrm{I_0}$ is the initial intensity of the transmitter, $G(\alpha_\mathrm{D}, \alpha_\mathrm{A})$ is calculated as (\ref{Attenuation}), with $\alpha_\mathrm{D}=\angle(\boldsymbol{r}_t, \boldsymbol{r}^{(1)})$ and $\alpha_\mathrm{D}=\angle(\boldsymbol{r}_r, \boldsymbol{r}^{(0)})$. Otherwise, if condition (\ref{condition}) is not satisfied, $\mathbf{I}(i,j)=0$.

\section{Proposed Vision-Based Beam Tracking Algorithm}\label{s3}
According to section \ref{s2}, the geometric angles of departure and arrival of the direct path keep changing due to the dynamic fluctuations of the water-air interface, which causes severe fluctuation of the receiving intensity. Thus, real-time transceiver steering/alignment becomes necessary to maintain reliable data transmission. 
To this end, we propose a novel beam alignment algorithm incorporating the visioned-based deep learning technology, which is capable of real-time environmental sensing and robust signal maintenance.


\subsection{Data Pre-processing}\label{s3.1}
The raw input data $\mathbf{I} \in \mathbb{R^*}^{\mathrm{M}\times\mathrm{N}}$ is captured by the photosensitive screen under the RT method mentioned in section \ref{s2}. Obviously, to fit the input form of the neural network, the raw data requires pre-processing. 
First, a clipping layer is applied to the raw data $\mathbf{I}$, clipping it from $M\times N$ to $n_{1x}\times n_{2x}$ vector $\mathbf{I}'$, where $n_{1x}$ and $n_{2x}$ represent the length and width of the input vector.
Second, the clipped vector $\mathbf{I}'$ is processed by a normalization step $\mathbf{x_i} = \frac{1}{\max{\mathbf{I}'}}\mathbf{I'}$, which converts the clipped vector into a normalized grayscale vector.
Finally, time-sequenced vectors are connected, forming the input data vector $\mathbf{X} = \{\mathbf{x}_1,\mathbf{x}_2,\ldots,\mathbf{x}_T\}$
The dataset applied by this training process is a set of time-sequenced vectors sampled from random environments, and labeled by prior-recognized directional vector $\mathbf{y} = [y_1,y_2,y_3]^\top$ indicating the direction of the optical communication link in 3-dimentional coordinate system. 
Before the training process, the dataset is divided into training, validation, and test sets in the ratio of $70\%$, $15\%$, $15\%$ with each training sample comprises consecutive frames in standard length.

\subsection{Vision-Intelligence based Beam Tracking Algorithm}\label{s3.2}
Based on the aforementioned system model, an AI-empowered optical beam-tracking algorithm is developed.
The proposed network architecture is constituted by CNN layers and bi-LSTM layers, which incorporates the attention mechanism to effectively track dynamic information across image sequences\cite{CNN_LSTM}. Balancing accuracy and computational complexity, the architecture is set with five convolution-pooling groups with increasing feature depth in CNN structures and $128\times 2$ LSTM units. The network processes input sequences of grayscale image set $\mathbf{X} = \{\mathbf{x}_1,\mathbf{x}_2,\ldots,\mathbf{x}_T\}$ which contains $n_t$ images in the size of $n_{1x}\times n_{2x}$ pixels, and outputs directional vector $\mathbf{y} = [y_1,y_2,y_3]^\top$ which contains directional information to guide beam alignment.
As shown in Fig. \ref{CNNLSTM}, the network comprises three main components: CNN extractor, bi-LSTM structure, and attention mechanism, which is expressed in (\ref{networkFunc.}). 
First, the CNN structure $\Theta_{\text{CNN}}(\cdot)$ extracts spatial features through convolutional layers, each of which connects convolution, batch normalization, ReLU activation, max pooling, and dropout layer. This progressively reduces the spatial dimensions of the vector $\mathbf{X}$ while increasing the feature depth. Afterwards, the output of the 2D ConvNet structure is further compressed by a dense layer connecting to the following layer.
Second, bidirectional-LSTM structure $\overrightarrow{\Theta_{\text{LSTM}}}(\cdot)$ and $\overleftarrow{\Theta_{\text{LSTM}}}(\cdot)$ is connected after the CNN network layers to capture sequential dependencies across dimensions of the vectors for temporal processing. The bidirectional design allows the network to consider both the past and future context, producing $M \times 2$-dimensional features for each time step. 
The attention mechanism then assigns importance weights $v^\top$ to different time features, generating a context vector that emphasizes the most relevant temporal information for prediction, which has been proved by experimental training to have higher accuracy than CNN-LSTM without the assistance of attention mechanism\cite{Action_recognition}.
At the final part of the network, a two-layer full connection network $\Theta_{\text{FC}}(\cdot)$ compresses the output of former networks and reshapes the context vector into a 2D output, linking to tanh activation layer that constrains the values between $[-1,1]$. 
Additionally, flexible parameters in (\ref{CNNLSTM}) such as: layer bias $b_o$ and $b_a$, and output weight adaptive matrix $W_o$ enable the adaption of the network.

\subsection{Network Training}\label{s3.3}

\begin{figure}[!h]
    \centering
    \includegraphics[width=1\linewidth, keepaspectratio]{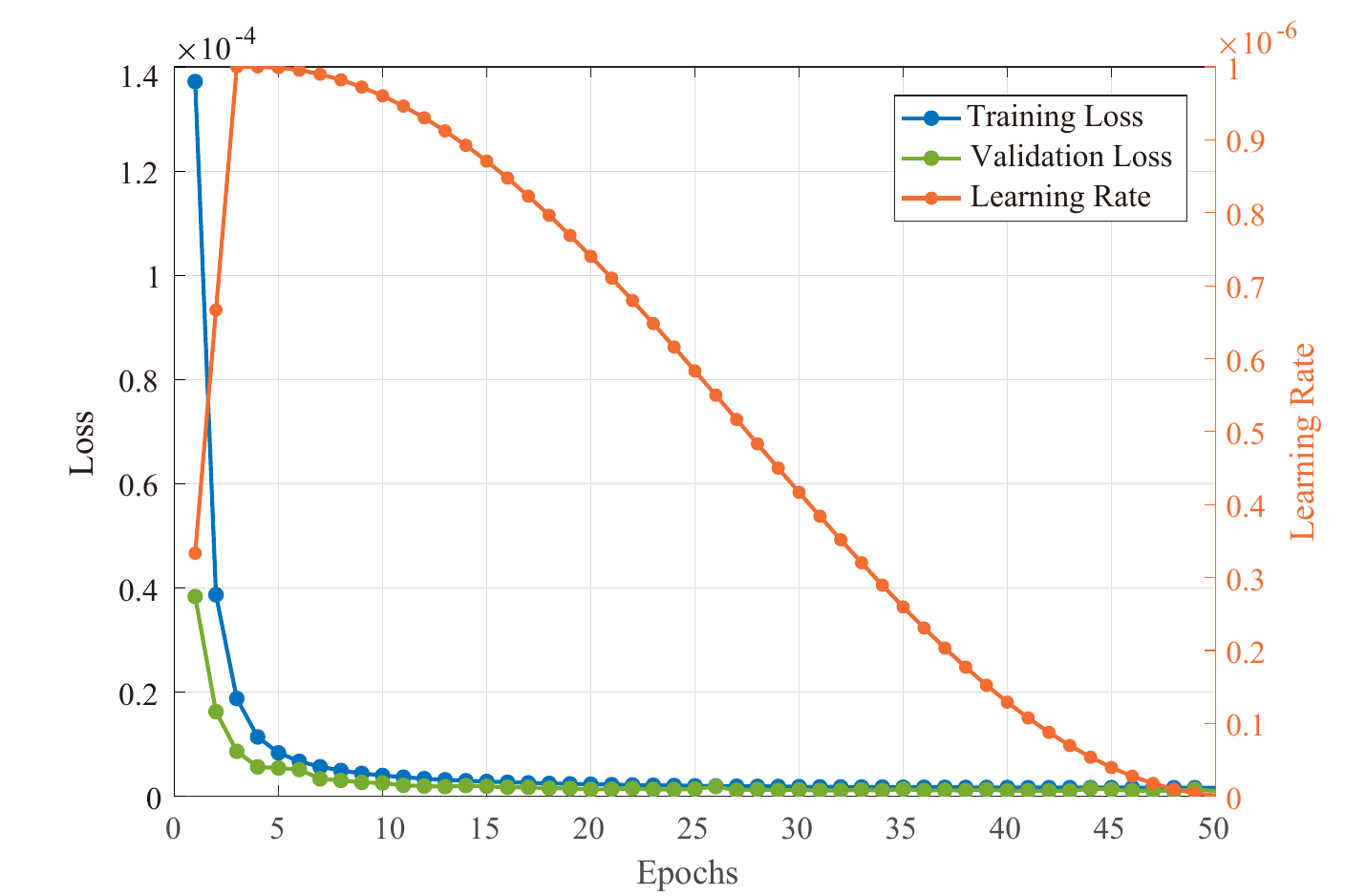}
    \caption{Training History, including training/validation loss and learning rate of the Proposed Algorithm in 50 epochs.}
    \label{Training_History}
\end{figure}

%
Based on the training data and network structure mentioned in section \ref{s3.1} and \ref{s3.2}, the proposed network is trained under the system model proposed in section \ref{s2}.
%
During each forward propagation of the training epoch, the proposed neural network maps input training data $\mathbf{X}$ to output vector representations $\hat{\mathbf{y}} = \hat{\Theta}({\mathbf{X}})$, where $\hat{\Theta}(\cdot)$ represents the network parameters.
Then, the back-propagation process handles gradient flow through attention, bi-LSTM, and CNN weights. 
A combined loss function shown in (\ref{Lossfunc}) is employed, which integrates mean squared error (MSE) and L1 loss with a ratio factor $\alpha$, leveraging the sensitivity of MSE to large deviations while benefiting from L1's robustness to small variations
\begin{equation}\label{Lossfunc}
    \begin{aligned}
        \mathcal{L}(\mathbf{y}, \hat{\mathbf{y}}) = \alpha \mathcal{L}_{\text{MSE}}(\mathbf{y}, \hat{\mathbf{y}}) + (1-\alpha) \mathcal{L}_{\text{L1}}(\mathbf{y}, \hat{\mathbf{y}}) \\
        = \alpha\cdot \frac{1}{2}||\mathbf{y} - \hat{\mathbf{y}}||_2^2 - (1-\alpha)\cdot ||\mathbf{y} - \hat{\mathbf{y}}||_1.
    \end{aligned}
\end{equation}
An adam optimizer with an adaptable learning rate and weight decay drives parameter updates, supplemented by a cosine annealing learning rate scheduler that gradually reduces the learning rate to $1\%$ of its initial value for stable convergence. 
Moreover, data augmentation techniques including random rotation ($\pm5^\circ$), minor translation ($\pm5$ pixels), and Gaussian noise $\sigma$ enhance model generalization, which increases the quality of the network. 
Consequently, the training process under the aforementioned settings is illustrated in Fig. \ref{Training_History}, in which the learning rate decreases to provide training accuracy, while the training and validation loss demonstrate the model's convergence during the training process.

\section{Numerical Results and Discussions}\label{s4}
The numerical results for the proposed algorithm are presented in this section. The performance of the proposed methods and a group of counterparts are tested under a simulated oceanic environment to validate their efficiency for alignment task.

\begin{table}[!b]
    \label{Parameter_Settings}
    \centering
    \caption{Environmental Simulation Parameters.}\centering
    \scalebox{0.9}
    {
        \begin{tabular}{|l|l|l|}
        \hline
            $\textbf{Parameter}$ & $\textbf{Symbol}$ & $\textbf{Value}$ \\ \hline
            \multicolumn{3}{|l|}{$\textbf{Environment}$} \\ \hline
            Gravity & g & 9.80 $\mathrm{m}/\mathrm{s}^2$ \\ \hline
            Wind Speed at 10m altitude & $U_{10}$ & 10 m/s \\ \hline
            Average Fetch & x & 2e4 m \\ \hline
            Spectrum shaping factor & $\gamma$ & 3.3 \\ \hline
            \multicolumn{3}{|l|}{$\textbf{Channel}$} \\ \hline
            Underwater Absorption Coefficient & $a_w$ & 1.80e-2 $\mathrm{m}^{-1}$ \\ \hline
            Underwater Scattering Coefficient & $b_w$ & 3.81e-3 $\mathrm{m}^{-1}$ \\ \hline
            Aerial Absorption Coefficient & $a_a$ & $\leq$ 1e-7 $\mathrm{m}^{-1}$ \\ \hline
            Aerial Scattering Coefficient & $b_a$ & 2.96e-5 $\mathrm{m}^{-1}$ \\ \hline
            Maximum Departure angle & $\omega_D$ & $\pi/600$ \\ \hline
            Maximum Arrival angle & $\omega_A$ & $\pi/3$ \\ \hline
            \multicolumn{3}{|l|}{$\textbf{Vision settings}$} \\ \hline
            Screen Sampling Rate & r & 60 fps \\ \hline
            Equivalent Focusing Length & f & 15 mm \\ \hline
            Pixel interval & $d_\mathrm{px}$ & 0.1 mm \\ \hline
        \end{tabular}
    }
\end{table}

\subsection{Simulation Environment Setup}\label{s4.1}
\begin{figure}[!t]
    \centering
    \includegraphics[width=0.9\linewidth, keepaspectratio]{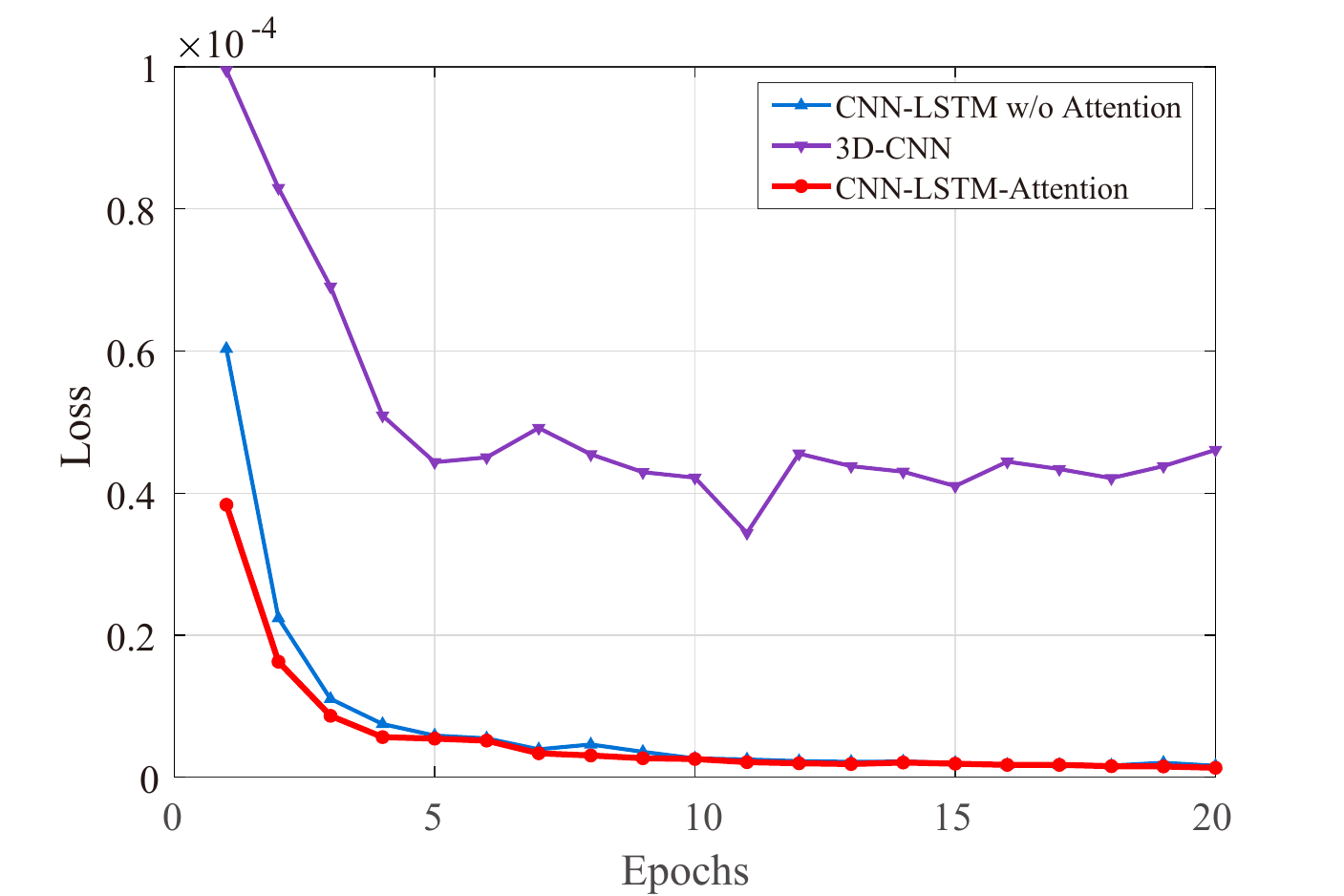}
    \caption{Testing result of experimental training between the proposed algorithm, CNN-LSTM w/o attention, and 3D Convolutional network.}
    \label{expTraining}
    \vspace{-3mm}
\end{figure}

\begin{figure}[!t]
    \centering
    \includegraphics[width=1\linewidth, keepaspectratio]{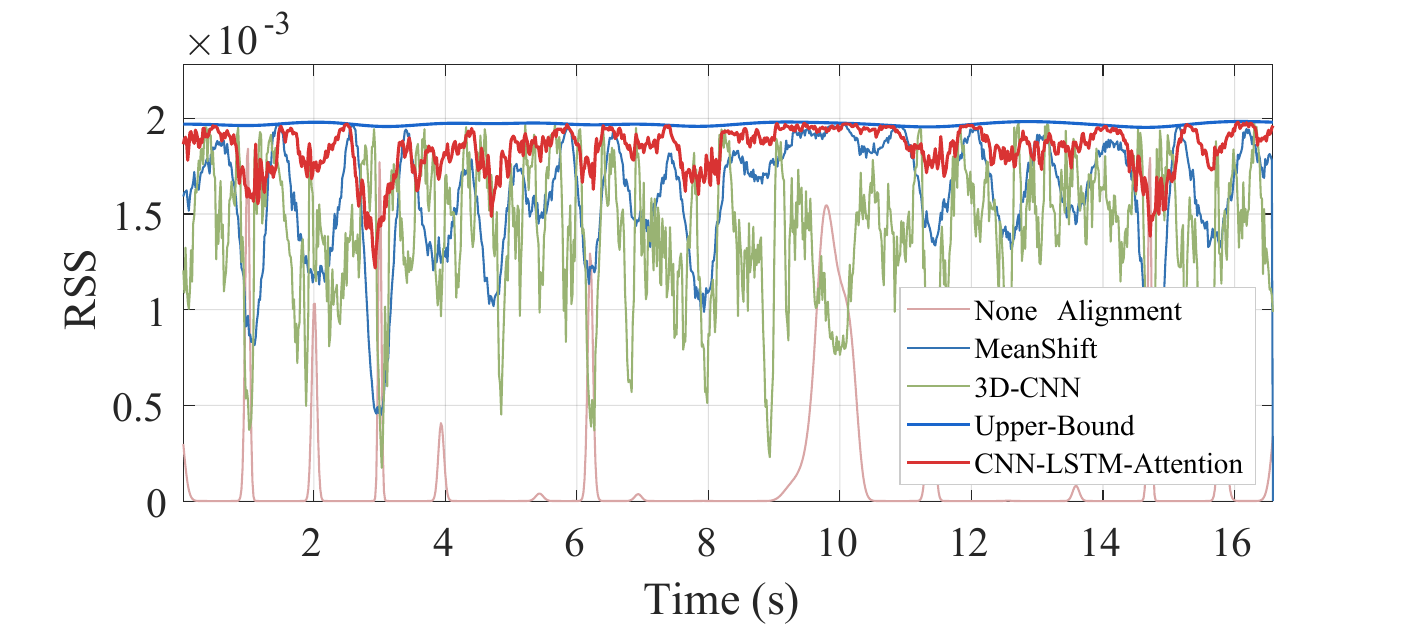}
    \vspace{-3mm}
    \caption{Received signal strength under different optical beam alignment algorithms.}
    \label{TemporalGain}
    \vspace{-3mm}
\end{figure}

A simulation environment, with its architecture illustrated in Fig. \ref{sysmodel}, is established based on the oceanic environment described in section (\ref{s2}), and the channel model formulated as (\ref{Channel_model}). 
Moreover, noise that influences the whole system is built, including relative intensity noise, environment noise (solar noise), thermal noise, and shot noise\cite{Background_Noise}.
To maintain experimental reproducibility and stability, the channel's stochastic components are generated using fixed random seeds which are different from seeds employed in the training dataset. The simulation framework comprises three primary components: environmental model, signal transmission model, and signal processing model. The parameter configurations for these components are shown in detail as the following TABLE I,
where the environmental parameters are mainly influential to the wave spectrum described as (\ref{WaveSpectrum}), the channel parameters decide the value of the channel attenuation shown in (\ref{Attenuation}), and the signal processing parameters are influential to the simulation of vision images which are input of the optical beam alignment network.

For performance validation, the proposed algorithm was compared with two state-of-the-art visual processing networks, specifically 3D-CNN and conventional CNN-LSTM architectures. Both networks are trained under the same dataset and with similar hyperparameters. 
The training record shown in the Fig. \ref{expTraining} indicates that both of the networks are convergence, while our proposed approach get lower validation loss than its counterparts.

\subsection{Results and discussions}\label{s4.2}

\begin{figure}[!t]
    \centering
    \includegraphics[width=0.9\linewidth, keepaspectratio]{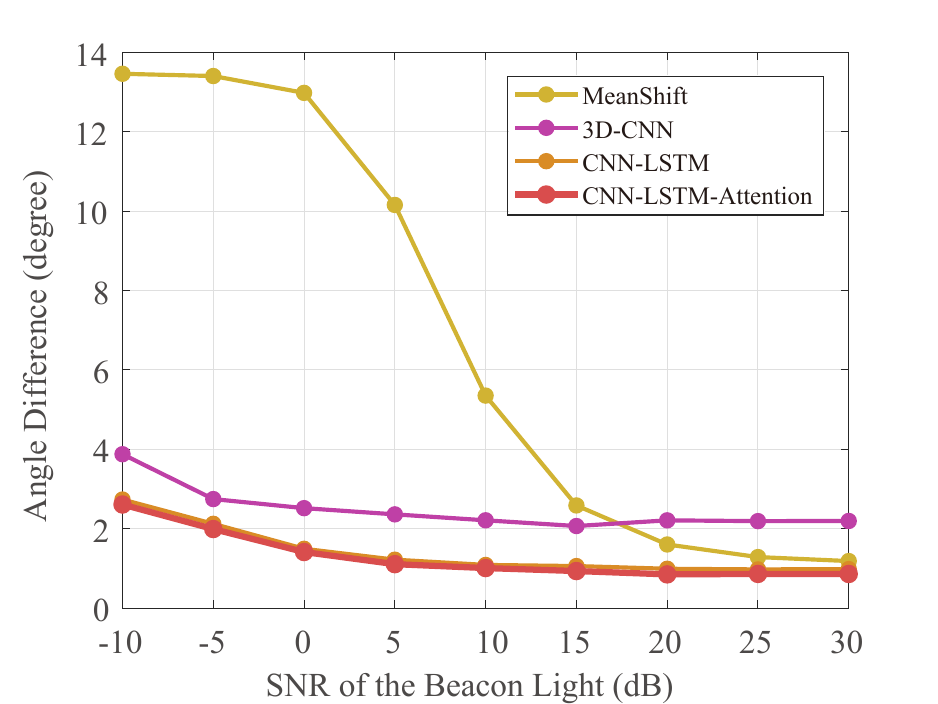}
    \caption{Average angle difference over vision noise.}
    \label{Avgang}
\end{figure}
\begin{figure}[!t]
    \centering
    \includegraphics[width=0.9\linewidth, keepaspectratio]{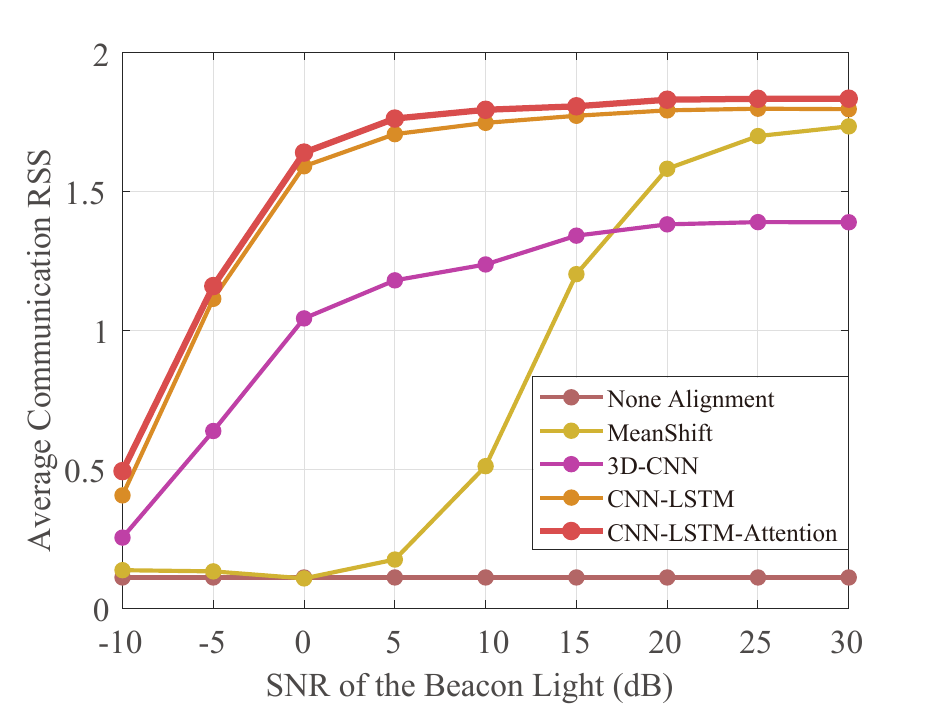}
    \caption{Average RSS over vision noise.}
    \label{Avgrss}
    \label{fig:placeholder}
\end{figure}
The temporal received signal strength (RSS) provides an intuitive perspective to the effectiveness of the beam alignment algorithms. Fig. \ref{TemporalGain} records the real-time intensity receiving curves over a continuous period with constant transmission intensity, comparing the proposed method with 3D-CNN and mean-shift algorithms. The proposed method achieved an average RSS performance $15.7\%$ and $32.5\%$ higher than mean shift, and 3D-CNN counterparts, respectively. Moreover, with normalized transmission intensity, the proposed algorithm shows an RSS variance of $1.4\times10^{-5}$, representing a considerable improvement over $9.96\times10^{-5}$ and $1.42\times10^{-4}$ for mean shift and 3D-CNN , respectively.
The aforementioned result highlights the enhanced capabilities of the proposed algorithm compared to its counterpart.


The superior performance of the proposed algorithm might be attributed to the visual scheme integrated with the algorithmic structure, which enables more precise sensing of the beam direction and aligns the transceiver orientation closer to the minimum OPL path. 
This configuration enhances the direct LOS path gain at the receiver while reducing interference.
Compared with traditional algorithms such as mean shift, the proposed algorithm possesses the capability to predict spatio-temporal information. In contrast to other AI methods, the proposed algorithm incorporates attention mechanisms that enable clearer discrimination of effective signal directions from noise.
The result in Fig. \ref{Avgang} demonstrates the inference for the proposed algorithm has lower angle difference with the path of minimum OPL, especially in strong environmental noises. 
The results in Fig. \ref{Avgrss} align with the aforementioned expectations, demonstrating that alignment methods with less angular deviations from the minimum OPL path achieve higher average gains, thereby further validating the above inference.

\begin{figure}[!t]
    \centering
    \includegraphics[width=1\linewidth, keepaspectratio]{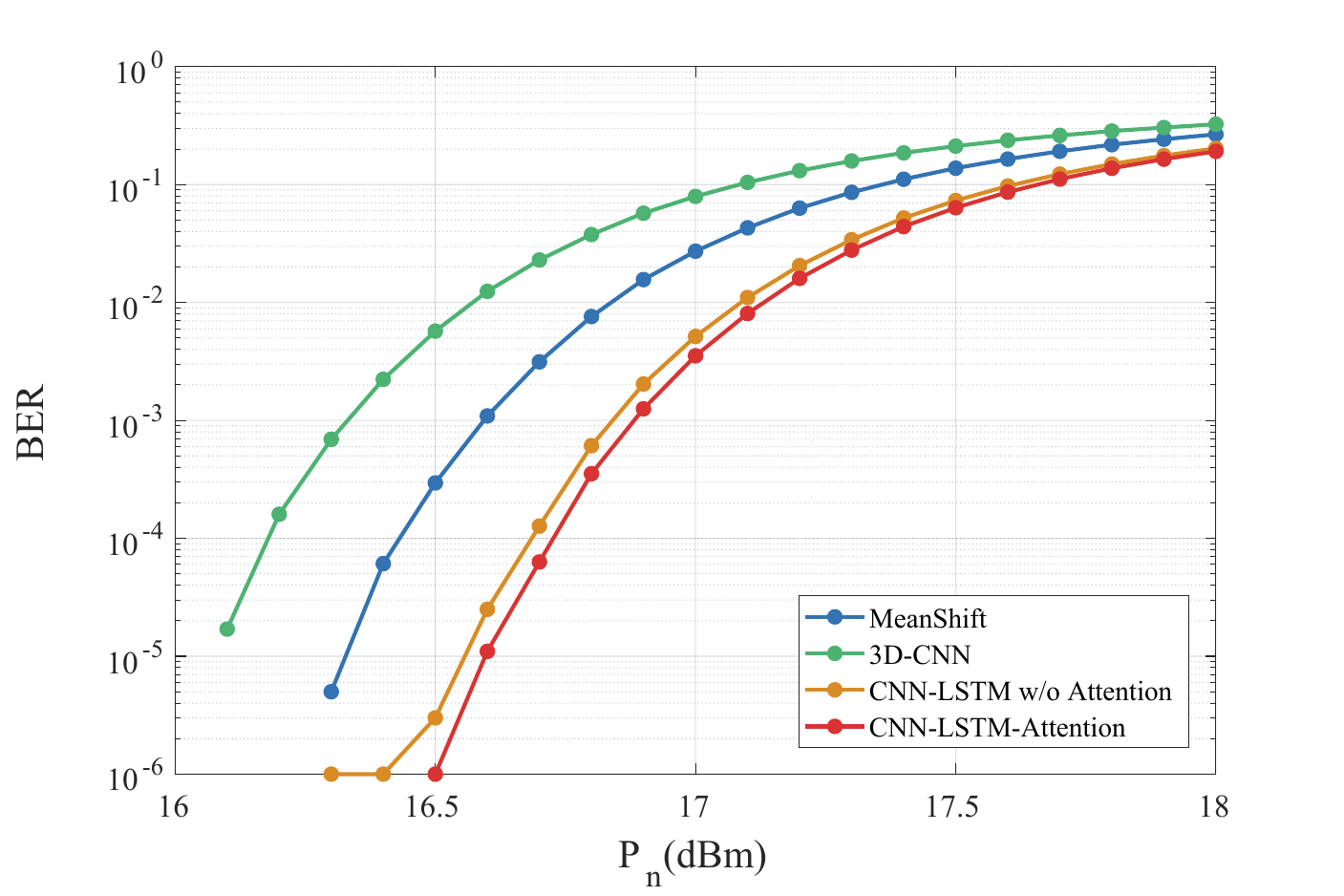}
    \caption{Simulated temporal communication BER.}
    \label{BER}
\end{figure}

Finally, from a holistic communication system perspective, simulation of the average bit error rate (BER) over noise is displayed in Fig. \ref{BER}, which describes the performance of the algorithm in the communication scenario. Here $P_n$ denotes the power of constant-intensity noise, i.e, thermal noise and solar noise. The result has further demonstrated the superiority of the proposed algorithm for lower BER over environmental noise. 
Consequently, the aforementioned result demonstrates that the proposed beam alignment algorithm shows superior performance in increasing alignment accuracy and reducing the influence of vision noise compared to the traditional mean shift and 3D-CNN benchmarks. 


\section{Conclusion}\label{s5}
This paper presents an analysis of water-air cross-interface OWC system between LAP and underwater platform, focusing on optical beam alignment tasks in dynamic oceanic environments. 
A condensed optical channel model is proposed to analyze propagation path and attenuation of the optical wireless channel.
Based on the system, optical beam alignment is achieved by an AI-based approach, utilizing an attention-enhanced CNN-LSTM architecture, whose spatio-temporal prediction and attention capabilities make it ideal for beam alignment task under complicated environment.
The result of the numerical simulation demonstrates that the proposed algorithm outperforms its counterparts in high alignment accuracy, stable signal strength maintenance, and considerable noise resilience against the dynamic water-air interface, reaching higher BER performance and better link stability.


\bibliographystyle{ieeetr}
\bibliography{Reference}
\vspace{-1mm}

\clearpage

\end{document}